\documentclass[letter]{IEEEtran}
\usepackage{amssymb}
\usepackage{amsmath,amsfonts,amssymb}
\usepackage[dvips]{graphicx}
\usepackage{verbatim}
\usepackage{slashbox}
\usepackage{bm}
\usepackage{algorithm} 
\usepackage{algorithmic} 
\usepackage{multirow} 
\usepackage{amsmath}
\usepackage{xcolor}

\usepackage[bookmarks=false]{}

\IEEEoverridecommandlockouts

\begin{document}

\title{Suboptimal Spatial Diversity Scheme for 60 GHz Millimeter-Wave WLAN}

\author{
Zhenyu Xiao,~\IEEEmembership{Member,~IEEE}
\thanks{This work was supported by the National Natural Science Foundation of China (NSFC) under grant No. 61201189, the Postdoctoral Science Foundation under grant No. 2011M500326 and 2012T50094.}
\thanks{The author is with the School of
Electronic and Information Engineering, Beihang University, Beijing 100191, P.R.
China.}
}

\maketitle

\begin{abstract}
This letter revisits the equal-gain (EG) spatial diversity technique, which was proposed to combat the human-induced shadowing for 60 GHz wireless local area network, under a more practical frequency-selective multi-input multi-output channel. Subsequently, a suboptimal spatial diversity scheme called maximal selection (MS) is proposed by tracing the shadowing process, owing to a considerably high data rate. Comparisons show that MS outperforms EG in terms of link margin and saves computation complexity.
\end{abstract}


\vspace{-0.2 in}
\section{Introduction}
\IEEEPARstart{T}{he emerging} IEEE 802.11ad wireless local area network (WLAN) standard promises multi-giga bits per second (Gbps) transmission by exploiting the 60 GHz communications \cite{Eldad 2010}-\cite{IEEE Std 802.11ad}, where beamforming technique is necessary to compensate for high path loss. Despite this, human-induced shadowing, especially blocking, may easily break a link due to the stringent link budget. To cope with this, a re-beamforming process can be initiated to find an alternative link \cite{An Xueli}, or a multihop scheme can be adopted to bypass the blockage through one or more relay nodes \cite{IEEE Std 802.11ad}, \cite{Singh}. These approaches, however, have the problem that the transmitter and receiver detect that the link is lost only after dropping a large amount of data due to the high data rate and a large packet size \cite{Minyoung Park 2012}.

To address this problem, a spatial diversity technique was proposed by Park and Pan in their recent work \cite{Minyoung Park 2012}, where multiple beams along the $N$ strongest multiple propagation paths are formed simultaneously during a beamforming process, so that when one of the propagation paths is blocked by a human, there are other propagation paths left to maintain the communication link. As the power gain on each path is set to be equal, the scheme is called equal-gain (EG) diversity scheme. Although a frequency-flat multi-input multi-output (MIMO) channel was adopted in their work, the EG scheme is proven effective to combat human-induced shadowing via simulation and experiments.

In this letter the EG scheme is revisited under a frequency-selective (FS) MIMO channel, which is more practical for 60 GHz communications, because the bandwidth is sufficiently large to resolve multipaths \cite{IEEE Std 802.11ad}.\footnote{Note that in \cite{IEEE Std 802.11ad} the OFDM sample time is 0.38 ns and the single-carrier (SC) sample/chip time is 0.57 ns, both of which are small enough to resolve multipaths.} Moreover, explicit expressions of total power gain are presented, which are not provided in \cite{Minyoung Park 2012} but necessary in computation of received power. Subsequently, realizing that transmitting a packet is much faster than human-induced shadowing for 60 GHz WLAN owing to the multi-Gbps speed, a suboptimal spatial diversity scheme called maximal selection (MS) is proposed by tracing the shadowing process. Comparison results on received power and bit-error rate (BER) show that the proposed scheme not only achieves a higher link margin in both normal and blocked cases, but also reduces implementation complexity.

\vspace{-0.1 in}
\section{Channel Model}
Let $N_t$ and $N_r$ denote the number of transmit and receive antennas, respectively. A MIMO FS channel model is adopted here. Following the conventions used in \cite{Minyoung Park 2012}, we assume that the $l$-th reflector is located in direction $(\phi_{tl},\theta_{tl})$ from the transmitter, and $(\phi_{rl},\theta_{rl})$ from the receiver. The transmit steering vectors, ${{\bf{h}}_l}$, corresponding to the $l$ reflector and associated with direction $(\phi_{tl},\theta_{tl})$, is expressed as ${{\bf{h}}_l}=\frac{1}{\sqrt{N_t}}[e^{j2\pi f_0\tau_1(\phi_{tl},\theta_{tl})},...,e^{j2\pi f_0\tau_{N_t}(\phi_{tl},\theta_{tl})}]^T$, where $f_0$ is the carrier frequency of the signal, $\tau_{1}(\phi_{tl},\theta_{tl})=0$ and $\tau_{i}(\phi_{tl},\theta_{tl})$ is the relative delay for the $i$-th transmit antenna versus the first transmit antenna to the same receive antenna over the $l$-th path, $(\cdot)^T$ is the transpose operator. Similarly, the receive steering vectors, ${{\bf{g}}_l}$, corresponding to the $l$-th reflector and associated with direction $(\phi_{rl},\theta_{rl})$, is expressed as ${{\bf{g}}_l}=\frac{1}{\sqrt{N_r}}[e^{j2\pi f_0\tau_1(\phi_{rl},\theta_{rl})},...,e^{j2\pi f_0\tau_{N_r}(\phi_{rl},\theta_{rl})}]^T$, where $\tau_{1}(\phi_{rl},\theta_{rl})=0$ and $\tau_{i}(\phi_{rl},\theta_{rl})$ is the relative delay for the $i$-th receive antenna versus the first receive antenna to the same transmit antenna over the $l$-th path. Thus, the channel matrix over the $l$-th path can be expressed as ${{\bf{C}}_l}={{\bf{g}}_l}\lambda_l {{\bf{h}}_l^T}$, where $\lambda_l$ is the channel coefficient of the $l$-th path. Subsequently, taking the multipath delay into account, the FS channel matrix is obtained as
$
{\bf{C}}[k] = \sum_{l = 1}^N {{{\bf{C}}_l}\delta [k - {\Delta _l}]}
$,
where $N$ is the number of multipaths, $\Delta_l$ is the normalized delay from the first transmit antenna to the first receive antenna over the $l$-th path. It is important to note that $\Delta_l$ was not involved in \cite{Minyoung Park 2012}; thus, the channel model reduced to a frequency-flat one there.

\vspace{-0.1 in}
\section{EG Diversity Revisit}
The $m$-th received sample $y[m]$ over the $N$ paths is expressed as \vspace{-0.1 in}
\begin{equation} \label{eq_ym}
\begin{aligned}
y[m] &= \sum_{l=1}^N{\bf{w}}_r^T{\bf{C}}_l\frac{{{\bf{w}}_t}}{{\sqrt {{\bf{w}}_t^H{{\bf{w}}_t}} }}s[m-\Delta_l] + {\bf{w}}_r^T{\bf{n}},
\end{aligned}
\end{equation}
where $s[m]$ is the $m$-th transmitted sample with an average power $P$, ${\bf{w}}_t$ and ${\bf{w}}_r$ are transmit and receive antenna weight vectors (AWVs), respectively, $\bf{n}$ is a circularly symmetric complex Gaussian noise vector with
identical variance for each element. Defining the transmit and receive antenna gain over the $l$-th path as $\alpha_l={\bf{h}}_l^T{\bf{w}}_t$ and $\beta_l={\bf{w}}_r^T{\bf{g}}_l$, respectively, we have
\vspace{-0.1 in}
\begin{equation}
y[m]= \frac{1}{{\sqrt {{\bf{w}}_t^H{{\bf{w}}_t}} }}\sum\limits_{l = 1}^N {{\alpha _l}{\beta _l}{\lambda _l}s[m-\Delta_l]}  +
{\bf{w}}_r^T{\bf{n}}.
\end{equation}
Let ${\lambda _l^{(0)}}$ denote the channel gain, which accounts for the effect of propagation loss and reflection loss of the $l$-th path
when beamforming is performed. As EG sets identical power gains, which are channel gains multiplied by antenna gains over each path, we achieve $
{\alpha _l}{\beta _l} = {{{{\bar \lambda }^{(0)}}}}/{{\lambda
_l^{(0)}}},
$
where ${{\bar \lambda }^{(0)}} = \frac{1}{N}\sum_{l = 1}^N {\lambda
_l^{(0)}}$. Hence $y[m]$ can be expressed as
\begin{equation} \label{eq_ym1ast}
y[m] = \frac{1}{{\sqrt {{\bf{w}}_t^H{{\bf{w}}_t}} }}\sum\limits_{l =
1}^N {\frac{{{{\bar \lambda }^{(0)}}}}{{\lambda _l^{(0)}}}{\lambda
_l}s[m-\Delta_l]}  + {\bf{w}}_r^T{\bf{n}}.
\end{equation}
It is noted that in a non-shadowing case, the channel gains do not vary, i.e.,
$\lambda_l={{\lambda_l }^{(0)}}$; thus, $y[m] = {1}/{{\sqrt
{{\bf{w}}_t^T{{\bf{w}}_t}} }}\sum_{l = 1}^N {{{\bar \lambda }^{(0)}}s[m-\Delta_l]} + {\bf{w}}_r^T{\bf{n}}$. However, in a shadowing case it
does not hold since $\lambda_l \neq {{\lambda_l }^{(0)}}$ once $\lambda_l$ varies.

Taking the number of transmit and receive antennas into account, one appropriate way to determine $\alpha_l$ and $\beta_l$ is to constrain $\alpha_l/\beta_l=N_t/N_r$. Thus,
$
\alpha_l=\sqrt{{{{{\bar \lambda }^{(0)}}}}/{{\lambda
_l^{(0)}}}{N_t}/{N_r}}~~\text{and}~~\beta_l=\sqrt{{{{{\bar \lambda }^{(0)}}}}/{{\lambda
_l^{(0)}}}{N_r}/{N_t}}.
$
With both the amplitude and phase controlled (APC), the transmit and receive AWVs are obtained as
\begin{equation} \label{eq_awvapc}
\mathbf{w}_t=(\mathbf{H}^T)^{-1}\bm{\alpha}~~\text{and}~~
\mathbf{w}_r=\bm{\beta}^T\mathbf{G}^{-1},
\end{equation}
where $\bm{\alpha}=[\alpha_1,...,\alpha_N]^T$, $\bm{\beta}=[\beta_1,...,\beta_N]^T$, $\mathbf{H}=[{\bf{h}}_1,...,{\bf{h}}_N]$, $\mathbf{G}=[{\bf{g}}_1,...,{\bf{g}}_N]$, $(\cdot)^{-1}$ is the pseudo-inverse operation.

The total power gain for the EG scheme with APC is
\begin{equation} \label{tgain}
G_{EG-APC} = {{\sum\limits_{l = 1}^N {{{\bigg{|} {\frac{{{{\bar \lambda
}^{(0)}}}}{{\lambda _l^{(0)}}}{\lambda _l}}\bigg{|}^2}}} }}\big{/}\left({{\left(
{{\bf{w}}_t^H{{\bf{w}}_t}} \right)\left( {{\bf{w}}_r^H{{\bf{w}}_r}}
\right)}}\right),
\end{equation}
which is the power gain observed after the receive antenna array
versus that before the transmit antenna array. Thus, it depends on
the AWVs in both ends and the channel gains $\lambda_l$, as
shown in (\ref{tgain}).

Note that in the above derivations, EG with APC is adopted. In the case
that only phase can be controlled (PC), the channel gains cannot be
precisely set. In such a case, ${\bf{w}}_t$ and ${\bf{w}}_r$ are
obtained by
\begin{equation} \label{eq_awvpc}
\mathbf{w}_t=\text{exp}(j\angle((\mathbf{H}^T)^{-1}\bm{\alpha}))
~\text{and}~
\mathbf{w}_r=\text{exp}(j\angle(\bm{\beta}^T\mathbf{G}^{-1})),
\end{equation}
respectively, where $\angle$ is the phase operation. Thus, $y[m]$ is expressed as (\ref{eq_ym}) instead of (\ref{eq_ym1ast}), and the corresponding total power gain is
\vspace{-0.1 in}
\begin{equation} \label{gain}
{G_{EG-PC}} = {{\sum\limits_{l = 1}^N {{{\bigg{|}
{{\bf{w}}_r^T{{\bf{C}}_l}{{\bf{w}}_t}} \bigg{|}^2}}} }}\big{/}\left({{\left(
{{\bf{w}}_t^H{{\bf{w}}_t}} \right)\left( {{\bf{w}}_r^H{{\bf{w}}_r}}
\right)}}\right).
\end{equation}

\section{Suboptimal Diversity Scheme}
The EGC scheme is efficient when the LOS path is blocked. In the normal case that the LOS path is not blocked, however, it is not optimal because larger antenna gains are set to poorer paths, i.e., transmit power is wasted on the NLOS paths. Moreover, FS effect is
generated and intensified due to the identical-energy multipath components. In this section, the optimal gain setting is simply analyzed under an ideal assumption. Based on it, the corresponding suboptimal diversity scheme is proposed.

With the 2-norm of the transmit and receive AWVs constrained to unity, according to (\ref{eq_ym}), the optimal AWVs to maximize receive signal-to-noise ratio (SNR) is achieved by
\vspace{-0.0 in}
\begin{equation} \label{eq_opeq}
[{\bf{w}}_t^{opt},{\bf{w}}_r^{opt}] = \arg \mathop {\max }_{{{\bf{w}}_{\bf{t}}}{\bf{,}}{{\bf{w}}_{\bf{r}}}} {\kern 1pt} {\kern 1pt} {\kern 1pt} {\kern 1pt} {\kern 1pt} \sum\nolimits_{l = 1}^N |{{\bf{w}}_r^T{{\bf{g}}_l}\lambda _l}{{\bf{h}}_l^T{{\bf{w}}_{\bf{t}}}}|^2.
\end{equation}
Thus, the optimal gain setting is $\alpha_l^{opt}={\bf{h}}_l^T{\bf{w}}_t^{opt}$ and $\beta_l^{opt}=({\bf{w}}_r^{opt})^T{\bf{g}}_l$. When $N=1$, the optimal solution is easily obtained as ${\bf{w}}_t^{opt}={\bf{h}}_1^*$ and ${\bf{w}}_r^{opt}={\bf{g}}_1^*$, where $(\cdot)^*$ is the conjugate operation. However, when $N>1$, i.e., multipath exists, the optimal solution of (\ref{eq_opeq}) is considerably difficult to obtain. Even though exploiting an iterative approach is able to achieve the optimal solution \cite{xia 2008}, it is not necessary to achieve the optimal BER performance, because inter-symbol interference due to multipath is not involved in (\ref{eq_opeq}).

To facilitate the analysis, it is natural to assume that the multiple reflection directions do not overlap with each other, i.e., ${\bf{h}}_l^\dagger{\bf{h}}_m=0$ and ${\bf{g}}_l^\dagger{\bf{g}}_m=0$ when $l\neq m$, where $(\cdot)^\dagger$ is the conjugate transpose operation. In fact, when $N_t$ and $N_r$ are large, the beamwidth of ${\bf{h}}_l$ and ${\bf{g}}_l$ are narrow, and do not overlap with ${\bf{h}}_m$ and ${\bf{g}}_m$, respectively. Thus, ${\bf{h}}_l^\dagger{\bf{h}}_m$ and ${\bf{g}}_l^\dagger{\bf{g}}_m$ will approximately equal 0.

Let yet ${\bf{h}}_l^T{\bf{w}}_t=\alpha_l$ and ${\bf{w}}_r^T{\bf{g}}_l=\beta_l$. Under this ideal assumption, we have $\sum_{l=1}^{N}\alpha_l^2=1$ and $\sum_{l=1}^{N}\beta_l^2=1$, due to the 2-norm constraint of AWVs. Let $\lambda_n^2=\max(\{\lambda_l^2\}|l=1,2,...,N)$, we achieve
\vspace{-0.0 in}
\[
\sum\limits_{l = 1}^N | {\bf{w}}_r^T{{\bf{g}}_l}{\lambda _l}{{\bf{h}}_l^T{{\bf{w}}_{\bf{t}}}}|^2\leq \lambda_n^2\sum\limits_{l = 1}^N\beta_l^2\alpha_l^2
\leq \lambda_n^2\sum\limits_{l = 1}^N\beta_l^2\sum\limits_{l = 1}^N\alpha_l^2=\lambda_n^2,
\]
where the equality holds when $\alpha_l=\beta_l=\delta[l-n]$. This is the optimal gain setting under the assumption, which suggests that the antenna arrays in both the transmitter and the receiver should beamform towards the direction of the strongest path. In such a case, the received power is larger, and the FS effect is less, which both contribute to improving the link margin.

Now the remaining question is how the antenna arrays can always beamform to the direction of the strongest path without dropping data. Realizing that a 60 GHz WLAN achieves a multi-Gbps speed, much faster compared with a shadowing process, we propose the \emph{shadowing tracing} algorithm for this purpose, which is described as Algorithm
\ref{alg:ms}.

\begin{algorithm}[tb]\caption{The MS Scheme with \emph{shadowing tracing}}\label{alg:ms}
\begin{algorithmic}
\STATE \textbf{1) Initialize:}
\begin{quote}
Perform beamforming. Sort the channel gains ($\lambda_l^{(0)}$) in a
descending order. Store them and their corresponding steering
vectors ($\mathbf{H}$ and $\mathbf{G}$).
\end{quote}

\STATE \textbf{2) Normal Communication:}
\begin{quote}
Set $k=1$. The transceiver beamforms to and communicate over the 1-st path direction,
which is usually the LOS path. During communication, the channel gain of the path ($\lambda_1$) is estimated for every packet. When the 1-st path is being blocked, the channel gain $\lambda_1$ will decrease sharply. Once $\lambda_1<\lambda_{2}^{(0)}$, go to \textbf{3)}.
\end{quote}

\STATE \textbf{3) Reselection:}
\begin{quote}
Set $k=k+1$. The transceiver change beamforming towards the $k$-th path according to the stored steering vector, and estimate the current channel gain $\lambda_k$. If
$\lambda_k<\lambda_{\min(k+1,N)}^{(0)}$, which means the current path is also blocked, repeat \textbf{3)} if $k<N$; go to \textbf{1)} to restart beamforming if $k=N$. Otherwise go to \textbf{4)}.
\end{quote}

\STATE \textbf{4) NLOS Communication:}
\begin{quote}
Communication is continued over the new-selected $k$-th path. The shadowing on the 1-st path is traced periodically, i.e., communication on the current path pauses with a period $T_P$, and the transceiver beamform to the 1-st path to test whether the block moves away. If the estimated channel gain $\lambda_1$ becomes larger than $\lambda_{k}^{(0)}$, which means that the block is moving away, go back to \textbf{2)}. Otherwise the transceiver beamforms toward the k-th path to continue NLOS communication. If the time for re-beamforming comes, or $\lambda_{k}$ decreases dramatically due to another block on the current path, go to \textbf{1)} for re-beamforming.
\end{quote}
\end{algorithmic}
\end{algorithm}

It is clear that for MS the whole shadowing process is traced.
Exploiting the \emph{shadowing tracing} approach, the transceiver
can rapidly change its beam towards the current strongest path
without dropping data or time-costly re-beamforming once the on-communication path is being
blocked. In \cite{Minyoung Park 2012} the typical shadowing duration
is 664 ms. The data octets of the current IEEE 802.11ad packet are
specified to be within the range of 0-262143 \cite{IEEE Std
802.11ad}. Hence, the maximal packet duration is only $262143\times
8/10^9\times 10^3=2.097$ ms if the transmit speed reaches 1 Gbps,
which means that the packet duration is significantly smaller than the
decay duration and the latter can be well traced. Therefore, the MS scheme
is applicable in practice. When
communicating on the $k$-th path, the weight vectors for MS are
\vspace{-0.1 in}
\begin{equation} \label{eq_ms}
{{\bf{w}}_{{t}}}={\bf{h}}_{k}^* ~~\text{and}~~
{{\bf{w}}_{{r}}}={\bf{g}}_{k}^*.
\end{equation}

In such a case the total power gain can be calculated from
(\ref{gain}), which is
\begin{equation}
G_{MS}=\frac{|{{\bf{w}}_{{r}}}^T{{\bf{C}}_{{k}}} {{\bf{w}}_{{t}}}|^2}{{\left(
{{\bf{w}}_t^H{{\bf{w}}_t}} \right)\left( {{\bf{w}}_r^H{{\bf{w}}_r}}
\right)}}\bigg{|}_{{{\bf{w}}_{{t}}}={\bf{h}}_{k}^*,
{{\bf{w}}_{{r}}}={\bf{g}}_{k}^*}=\lambda_k^2N_tN_r,
\end{equation}
where $N_t$ and
$N_r$ are gains of the transmit and receive antenna arrays,
respectively.

It can be observed that, compared with the EG scheme, the
superior points of MS are (i) it has a lower computation complexity,
because there are no matrix inversions and multiplications when
calculating the weight vectors; (ii) it achieves a higher total power
gain and does not induce FS effect, because the antenna arrays always
beamform towards the direction of the current strongest path.

The extra cost of MS is \emph{shadowing tracing}. The channel gain
of the 1-st path needs to be estimated each packet in the
Normal Communication state, and with an appropriate period $T_P$
in the NLOS Communication state. As a channel
estimation sequence is defined in the standard IEEE 802.11ad frame
format \cite{IEEE Std 802.11ad}, \emph{shadowing tracing} in the
Normal Communication state does not cause additional
cost. However, \emph{shadowing tracing} in the NLOS
Communication state will degrade efficiency, because communication
needs a periodical temporary pause, and antenna arrays in both ends need to
change beamforming between toward the 1-st and the on-communication
path directions, which may elapse tens or hundreds ${\mu}$s. The efficiency degradation is $\eta=2T_{BS}/T_P$, where $T_{BS}$ is the beam-switching time, and $T_P$ is the estimation period, which should be significantly smaller than the shadowing duration. As the shadowing duration is about several hundred ms \cite{Minyoung Park 2012}, by selecting a relatively large estimation period, e.g., 20 ms, and a common beam-switching time, e.g., 100 ${\mu}$s, the typical
degradation of efficiency is only $0.2/20=1\%$, which is minimal and acceptable.


\section{Performance Evaluation}

\begin{figure}[t]
\begin{center}
  \includegraphics[width=8.3 cm]{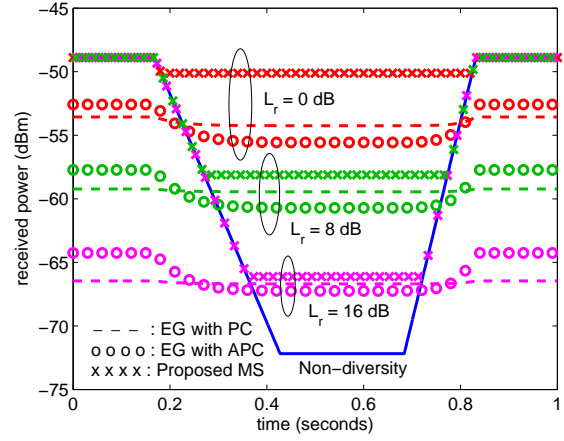}\vspace{-0.1 in}
  \caption{Comparison of received signal powers between EG, MS and the non-diversity scheme with different reflection losses. PC denotes phase control only, while APC denotes both amplitude and phase control. The drop of the received power is caused by human-induced shadowing.}
  \label{fig:results}\vspace{-0.2 in}
\end{center}
\end{figure}

Received powers are calculated with the same antenna placement,
human-induced shadowing model,\footnote{The shadowing duration was 664 msec, the decay time was 55.7 msec, the maximum attenuation was 23.3 dB, and the rise time was 31.8 msec.} path loss model, transmit power as
that in \cite{Minyoung Park 2012}. The reflection loss ($L_r$) is
set 0 to 16 dB in 8 dB step. A $20\times 1$ antenna array is used in
both the transmitter and the receiver. The received signal power
is achieved by adding the transmit power ($P=10$ dBm) and the corresponding
total power gain in dB.

Fig. \ref{fig:results} depicts the received signal powers for EG, MS
and the non-diversity scheme with different reflection losses. From this figure we observe that, as we expected, when
the LOS path is not blocked, EG receives a lower power than the
non-diversity scheme. EG with PC loses more power compared to that
with APC, whereas when the LOS path is blocked, EG with PC receives
a higher power than that with APC. By contrast, the proposed MS scheme
receives a higher power than the EG scheme in both non-blocked and
blocked cases. In the non-blocked case the superiority is more
evident when the reflection loss is larger; while in the blocked
case it is the opposite. We stress that the MS scheme has no power
loss compared with the non-diversity scheme when the LOS path is not
blocked.

\begin{figure}[t]
\begin{center}
  \includegraphics[width=8.2 cm]{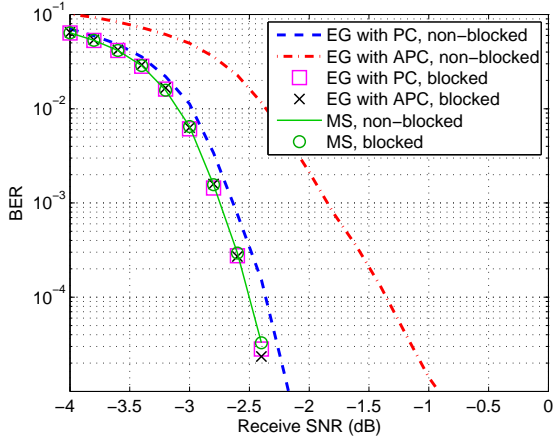}\vspace{-0.1 in}
  \caption{BER performance of EG and MS with $L_r=8$ dB in both blocked and non-blocked cases. The receive SNR is set the same in the blocked and non-blocked case to reflect the FS effect more clearly.}
  \label{fig:BER}\vspace{-0.2 in}
\end{center}
\end{figure}

In addition to the received power, the BER performance is also
evaluated via simulation, where carrier and timing synchronization,
as well as channel estimation, are assumed perfect. As the BER comparison here is to evaluate the FS effect, the receive SNR is set the same for all the cases.

The modulation and coding scheme 1 (MCS1) of SC PHY in \cite{IEEE Std 802.11ad}
with a chip time of $T_c=0.57$ ns is adopted and the SC frequency-domain equalization (SC-FDE) is used in the receiver to combat the FS effect. The typical reflection loss, i.e., $L_r=8$ dB, is exploited. For EG, there are two multipath components.
The relative delay for the NLOS path bounced by the ceiling versus the LOS path is
$[(\sqrt{(7/2)^2+2^2}\times 2-7)/(3\times 10^8)/0.57\times 10^{-9}]=6$
chip intervals,\footnote{According to the model in \cite{Minyoung Park 2012}, the LOS distance between the transceiver is 7m, and the height from the antennas to the ceiling is 2m. The propagation speed of 60 GHz signal is $3\times 10^8$ m/s.} where $[\cdot]$ is integer round operation. The
gains for the LOS and NLOS path are
$h_1=\lambda_1{{\bf{w}}_{{r}}^T}{\bf{g}}_{1}{\bf{h}}_{1}^T{{\bf{w}}_{{t}}}$
and
$h_2=\lambda_2{{\bf{w}}_{{r}}^T}{\bf{g}}_{2}{\bf{h}}_{2}^T{{\bf{w}}_{{t}}}$,
respectively, where ${\bf{h}}_{l}$ and ${\bf{g}}_{l}$ are determined by the antenna placement, $\lambda_1$ and $\lambda_2$ are computed according to propagation loss and reflection loss, ${\bf{w}}_t$ and ${\bf{w}}_r$ are calculated according to (\ref{eq_awvapc}) for APC and (\ref{eq_awvpc}) for PC. The equivalent normalized baseband channel response is $(h_1,0,0,0,0,0,h_2e^{-j2\pi
f6T_c})^T/\sqrt{|h_1|^2+|h_2|^2}$, where $f$ is the carrier
frequency and $f=60$ GHz. Note that the channel responses are
different between blocked and non-blocked cases, because
$\lambda_1$, the channel gain of the LOS path, varies. For MS, the channel response is similarly set.

The BER performance is shown in Fig. \ref{fig:BER}. It can be
observed that in the non-blocked case, EG with APC has a significant
loss compared with MS, while EG with PC has a smaller
loss. As SNR becomes larger, the gap between
EG with APC and MS becomes larger. Similar results can be observed with different parameter settings, e.g., transceiver distance, height of antenna, etc. This is because in the
non-blocked case EG with APC leads to two identical-energy
multipath components, which strengthens the FS effect. EG with PC
cannot strictly satisfy the target of equal gain on each path due to
the phase operation. Hence, the two multipath components have actually
different energy, which weakens the FS effect, and thus the corresponding BER is significantly better than that of EG with APC.
In the blocked case, most channel energy of EG with PC and APC
disperse on the NLOS path, because it has a larger channel gain and
antenna gain than the blocked LOS path. Consequently, the FS effect is little and the BER performance of EG with PC and APC become close to that of MS. Moreover, as MS always beamforms to the direction of the stronger path, the FS effect is little.

The overall link margin performance depends on both the received power and BER performance. If we jointly consider Fig. \ref{fig:results} and Fig. \ref{fig:BER} in
the case of $L_r=8$ dB, in the blocked case, compared
to EG with PC, MS achieves about a 10.3 dB higher receive power and a 0.1
dB SNR gain at $10^{-5}$ BER, i.e., a 10.4 dB higher link margin;
compared to EG with APC, MS achieves about an 8.8 dB higher receive
power and a 1.3 dB SNR gain at $10^{-5}$ BER, i.e., a 10.1 dB higher
link margin. In the non-blocked case, MS has no SNR gain
according to Fig. \ref{fig:BER}, but yet receives respectively 1.3 and 2.6 dB
higher link margin compared to EG with PC and APC due to the higher received power according
to Fig. \ref{fig:results}. In summary, MS achieves a higher link margin than EG with PA and APC in both cases, and the superiority is more significant in the non-blocked case.

\section{Conclusion}
The EG scheme has been revisited under a frequency-selective multipath MIMO channel for 60 GHz communications, and the total power gain that is necessary in the computation of received power has been obtained. Subsequently, the suboptimal MS diversity scheme has been proposed by exploiting the \emph{shadowing tracing} approach, which exploits the multi-Gbps speed of 60 GHz WLAN. Comparisons on the received power and BER show that MS has lower computation complexity, and achieves a higher link margin than EG, owing to the higher receive power and less FS effect. The superiority on link margin is more significant in the normal case, i.e., when the LOS path is not blocked.



\begin{thebibliography}{1}

\bibitem{Eldad 2010}
E. Perahia, C. Cordeiro, M. Park, and L. L. Yang,
``IEEE 802.11ad: Defining the next generation multi-Gbps Wi-Fi,''
in \emph{Proc. IEEE CCNC}, Jan. 2010, pp. 1-5.

\bibitem{M. Park 2008}
M. Park, C. Cordeiro, E. Perahia, and L. L. Yang, ``Millimeter-wave multi-gigabit WLAN: challenges and feasibility,'' in \emph{Proc. IEEE PIMRC}, Cannes, France, Sept. 2008, pp. 1-5.


\bibitem{IEEE Std 802.11ad}
\emph{Part 11: wireless LAN medium access control
(MAC) and physical layer (PHY) specifications -- amendment 3:
enhancements for very high throughput in the 60 GHz band}, IEEE P802.11ad$^\text{TM}$/D9.0, July 2012.

\bibitem{An Xueli}
 X. An, C. Sum, R. V. Prasad, etc., ``Beam switching support to resolve link-blockage problem in 60 GHz WPANs.'' In \emph{Proc. IEEE PIMRC}, Sept. 2009, pp. 390-394.

\bibitem{Singh}
S. Sumit, F. Ziliotto, U. Madhow, etc., ``Blockage and directivity in 60 GHz wireless personal area networks: from cross-layer model to multihop MAC design.'' \emph{Selected Areas in Communications, IEEE Journal on}, vol. 27, no. 8, pp. 1400-1413, Oct. 2009.

\bibitem{Minyoung Park 2012}
M. Park, H. K. Pan, ``A spatial diversity technique for
IEEE 802.11ad WLAN in 60 GHz band,'' \emph{IEEE Communications
Letters}, pp. 1260-1262, Aug. 2012.

\bibitem{xia 2008}
P. Xia, H. Niu, J. Oh, C. Ngo, ``Practical antenna training for millimeter wave MIMO communication,'' in \emph{Proc. IEEE VTC}, Sept. 2008, pp. 1-5.

\end{thebibliography}
\end{document}